\documentclass[prl, twocolumn]{revtex4}
\usepackage{graphicx}
\usepackage{amssymb}
\usepackage{epstopdf}
\begin{document}
A comment by Zhu (CZ) on our Letter \cite{plunk-prl} has recently been made \cite{zhu-comment}.  CZ raises several objections which we will address in this response.

\begin{itemize}
\item CZ proposes a hypothetical scenario involving a single ``frozen'' diagonal mode that mediates transfer among non-diagonal components.  It is concluded that this scenario is counter to our statement that ``transitions involving only one diagonal component are forbidden.''  To clarify our meaning, we employ the term ``transition'' to mean non-zero change, excluding this ``frozen'' scenario.
\item CZ states that the quasi-neutrality constraint, Eqn.~(3) of our Letter, is invalid due to the $k$-dependence of the velocity cutoff $v_{\mathrm{cut}}$.  This is incorrect.  A simple substitution of Eqn.~(1) into the quasi-neutrality equation, $\hat{\varphi} = \beta \int_0^{\infty}vdvJ_0(k v)\hat{g}$, yields precisely Eqn.~(3) of our Letter.  Although the derivation of Eqns. (3)-(6) is not completely explicit in our Letter, we stress that these equations have been derived without any approximation beyond that already stated in our Letter.  We plan to provide a more detailed presentation in a future publication.
\item A calculation of absolute equilibrium is provided by CZ, using a spectral representation similar to that provided in our Letter.  As this is original work by the author, we will make no comment on the calculation in this space but to say that we find no contradiction between it and our assumptions or conclusions.
\item CZ states that Eqns.~(1), (5) and (6) have been taken directly from expressions given in the continuum limit in a previous work \cite{plunk-jfm}.  This is not true.  These expressions have counterparts in the continuum limit, but have been derived separately for the discrete spectral representation in our Letter.
\item In reference to a normalizing constant that appears in the Bessel series expansion, CZ states ``the Letter replaces $J_1^2(x)$ with $x$.''  This is incorrect.  As stated in the Letter, we are concerned with the sub-Larmor limit ($k \gg 1$).  Although it was not explained explicitly in our Letter, we have used the large-argument form of the Bessel functions to evaluating $J_1$ at the zeros of the Bessel function, which results in a simple form of the Bessel series.
\item CZ makes the assertion that the constraints of our Letter ``do not predict the transfer and/or cascade directions'' and further comments that ``the arrows in Fig. 1 of the Letter could be reversed simultaneously!''  With regard to the transfer directions indicated in Fig.~(1), we clearly state in our Letter that ``the reverse process can also spontaneously occur.''  We later argue that the direction of nonlinear transfer can be predicted from a combination of the Fj{\o}rtoft argument and the {\it conjecture} that the redistribution of free energy is ``diffusive in $k$-$p$ space'' (this conjecture is supported by extensive numerical evidence).  We stress that the conclusions we draw from theoretical arguments are all supported by detailed numerical experiment.
\item CZ concludes that that our argument and the absolute equilibrium argument for dual cascade ``should not be thought to be independent, as [the Letter] indicated, for dual cascade which needs the combination of both of them, among other arguments.''  In our Letter we do not state that our argument is independent from the absolute equilibrium calculation.  We state in a footnote that it gives ``an alternative perspective for studying energy transfer'' and ``must be consistent.''  Although our work stands on its own, being a combination of arguments and numerical evidence, we do believe that other approaches are valid and valuable.
\end{itemize}

The above remarks are made in brief form to address the concerns raised by CZ.  We direct interested readers to an upcoming publication by the authors of \cite{plunk-prl}, which will provide more detailed explanations that should clarify both technical concerns and deeper issues of physical interpretation.
\\\\
G. G. Plunk and T. Tatsuno


\end{document}